\documentclass[journal]{IEEEtran}
\usepackage{cite}

\usepackage{svg}
\usepackage{mwe}
\usepackage{pgfplots}
\usepackage{subcaption}
\usepackage{circuitikz}
\usepackage{overpic}
\definecolor{lukas}{rgb}{0.1, 0.32, 0.12}
\definecolor{revised}{rgb}{0,0,0}
\ifCLASSINFOpdf
\else
\fi
\usepackage{todonotes}

\usepackage{amsmath}
\usepackage{siunitx}
\DeclareSIUnit{\fps}{fps}

\usepackage{booktabs}
\usepackage{tabularx}

\begin{document}
%
\title{Duality between Coronavirus Transmission and Air-based Macroscopic Molecular Communication }
%
%
%

\author{Max~Schurwanz,
        Peter Adam~Hoeher, 
        Sunasheer~Bhattacharjee,
        Martin~Damrath,
        Lukas~Stratmann, 
        and~Falko~Dressler
\thanks{M.~Schurwanz, P.\,A.~Hoeher, S.~Bhattacharjee, and M.~Damrath are with the Faculty of Engineering, Kiel University, Kiel, Germany, e-mail: \{masc,ph,sub,md\}@tf.uni-kiel.de.}
\thanks{L.~Stratmann and F.~Dressler are with the School of Electrical Engineering and Computer Science, TU Berlin, Berlin, Germany, e-mail: \{stratmann,dressler\}@ccs-labs.org.}
}
\maketitle

\begin{abstract}

This contribution exploits the duality between a viral infection process and macroscopic air-based molecular communication.
Airborne aerosol and droplet transmission through human respiratory processes is modeled as an instance of a multiuser molecular communication scenario employing respiratory-event-driven molecular {\color{revised} variable-concentration} shift keying.
{\color{revised}Modeling is aided by experiments that are motivated by a macroscopic air-based molecular communication testbed.
In artificially induced coughs, a saturated aqueous solution containing a fluorescent dye mixed with saliva is released by an adult test person. 
The emitted particles are made visible by means of optical detection exploiting the fluorescent dye. 
The number of particles recorded is significantly higher in test series without mouth and nose protection than in those with a well-fitting medical mask.}
A {\color{revised} simulation tool for} macroscopic molecular communication {\color{revised} processes} is extended and used for estimating the transmission of infectious aerosols in different environments. 
{\color{revised}Towards this goal, parameters obtained through self experiments are taken.}
The work is inspired by the recent outbreak of the coronavirus pandemic.
\end{abstract}

%
\IEEEpeerreviewmaketitle

\section{Introduction}
\IEEEPARstart{T}{he} recent outbreak of the \textit{severe acute respiratory syndrome coronavirus 2} (SARS-CoV-2) has led to an increased interest in the spread of infectious aerosols, generated by respiratory events of humans. Airborne aerosols are able to carry pathogens, for example in the form of viral ribonucleic acid (RNA), that might infect other humans when inhaled into the respiratory tract{\color{revised}~\cite{Bourouiba2014,Cann2005}}. An important factor in assessing the risk of infection is a good understanding of the propagation behaviour of aerosols in the air. Recent research corresponding to the coronavirus outbreak has largely dealt with the distribution of different-sized aerosols in the human airways~\cite{Madas2020} and the distribution of aerosols following air drifts in defined environments~\cite{Jayaweera2020} to assess different preventive techniques to control infection. Several publications have investigated the size distributions of aerosols in respiratory events like breathing and talking~\cite{Asadi2019,Anfinrud2020} as well as coughing and sneezing~\cite{Nishimura2013,Han2013,Tang2013}. Other parameters like the emission angle and initial velocity of particles have also been determined~\cite{Kwon2012}, ruling the distance that aerosols can travel~\cite{Xie2007}. Several studies have been conducted regarding the stability and concentrations of pathogenic aerosols on various surfaces~\cite{Doremalen2020,Guo2020}.
{\color{revised}Further research was conducted to determine the viral load in collected samples of particles from surfaces~\cite{Pan2020,Rothe2020,To2020}, as well as the influence of varying activity loads on the human body by the emitted viral load~\cite{Buonanno2020}.}

The literature uses the term ``aerosol'' for a wide range of particles, starting with aerosols of a few hundred nanometers in diameter up to droplets with a diameter of about $\SI{100}{\micro\meter}${\color{revised}~\cite{Hinds1999}}. In this paper, we follow the same approach as the literature for the descriptions of the scenario. In the evaluation of the measurement results, a distinction is made between droplets and cloud-like aerosols. When a distinction is not needed, however, reference is made to aerosols in general throughout the article.

The viral infection spread through aerosols and droplets can be seen from the perspective of molecular communication (MC). MC is often described as a nature-inspired approach to information transfer~\cite{farsad2016survey,haselmayr2019integration}. Molecules are used as carriers to transmit information between a transmitter and a receiver. From an evolutionary perspective, this concept has proven to be effective
{\color{revised}and can be observed in the communication of unicellular organisms via molecules in fluid environments~\cite{Bassler2002},}
or in macroscopic communication through pheromones in the air~\cite{Bassler2006}.
Starting from the microscopic domain with application areas inside the human body like targeted drug delivery~\cite{Akyildiz2010}, improved diagnosis and treatment with lab-on-a-chip devices to accelerate and refine the medical procedure~\cite{Nakano2013,Akyildiz2019}, in the domain of bio-sensors and intelligent materials~\cite{Nakano2005}, and even beyond in-body applications~\cite{dressler2015connecting}, MC is believed to improve a wide range of applications. 
{\color{revised}In addition to the theoretical discussions of MC application areas, there is currently a lack of practical implementations, as this requires a high degree of cooperation between engineers from different disciplines~\cite{Jamali2016}.} The macroscopic domain of MC mainly focuses on industrial and smart city applications for instance, where traditional radio communication might be challenging to accomplish. In order to understand the underlying principles and the overall system better, several testbeds have been developed~\cite{Jamali2019} based for example on magnetic nanoparticles~\cite{Unterweger2018}, alcohol molecules~\cite{Farsad2013}, or fluorescent dyes~\cite{Atthanayake2018, Bhattacharjee2020a, Bhattacharjee2020b} using a fluid-based or air-based environment.

First foundations of aerosol transmission from a communications point of view have recently been established in~\cite{Khalid2019,Khalid2020}. In \cite{Khalid2019}, a general concept is proposed and the duality between MC and aerosol transmission is highlighted. The transmission behavior of airborne pathogens is modeled by a macroscopic air-based MC scenario consisting of a transmitter, a distorted channel, and a receiver. This communication scenario has been described by Shannon in his landmark paper \cite{Shannon1948}, nowadays dubbed ``mother of all models.'' In the context of interest, the molecular transmitter is represented by an infected human, emitting virus-bearing aerosols into the environment. Different respiratory events like breathing, speaking, coughing, or sneezing define the properties of the aerosol emission process. 
{\color{revised}These include the amount of particles released with a certain probability distribution of different particle sizes, initial velocities, as well as emission directions.}
The random movement of the particles is described by the channel model. In \cite{Khalid2019}, a biosensor is assumed at the receiver side. The focus is on outdoor environments. In the follow-up work \cite{Khalid2020}, a complete system model is presented and an end-to-end mathematical model is derived for the transmission channel under certain constraints and boundary conditions. Additionally, the system response for both continuous sources such as breathing and jet or impulsive sources such as coughing and sneezing is studied. In addition to transmitter and channel, a receiver architecture composed of air sampler and silicon nanowire field-effect transistor is assumed. A detection problem to {\color{revised}maximize the decision likelihood} is formulated and the corresponding missed detection probability is minimized. Several numerical results are presented to observe the impact of parameters that affect the performance and justify the feasibility of the proposed setup in related applications.

Even more recently and independent of the presented work, the biosensor-based detector is replaced by an absorbing detector in the form of a healthy human located at a certain distance from the emitter~\cite{Gulec2020}. That way, the spreading mechanism of infectious diseases by airborne pathogen transmission between two humans could be modeled for the first time using the principles of MC. If a certain amount of aerosols is able to reach the receiver's detection space, the receiver is expected to be infected. This corresponds to a successful transmission of information between the transmitter and the receiver via a distorted channel in traditional MC. An end-to-end system model is proposed with emphasis on the indoor environment, predicting the infection state of a human. The corresponding infection probability is derived. 

Contrary to the classical data transmission approach, epidemiologists aim to reduce the number of successful transmissions.
In other words, in data transmission, a maximization of the mutual information is targeted, whereas in the case of virus transmission, a minimization of the mutual information is targeted. Based on the modeling assumptions, rules concerning ``social distancing'' are suggested. In the state-of-the-art \cite{Khalid2019,Khalid2020,Gulec2020}, the MC concept has been restricted to peer-to-peer links.

Our novel contributions include the following aspects:
\begin{itemize}
	\item The transmission of infectious {\color{revised} and non-infectious} aerosols is {\color{revised} interpreted as} a multiuser MC scenario. Transmitters are already infected humans and receivers are uninfected humans. Newly infected humans become new transmitters after some time -- the root of reproduction.
	\item In each MC transmitter, the modulator is modeled by respiratory-event-driven molecular variable-concentration shift keying {\color{revised}inspired by}~\cite{Bhattacharjee2019},  where the type of molecule {\color{revised}is used to distinguish between} coronavirus aerosols and any other particles (like noninfectious saliva). The number of released molecules per time unit depends on the situation: In the case of coughing, for example, more molecules are emitted than in the case of breathing.
	\item Channel modeling is aided by human experiments using a {\color{revised}fluorescent} dye in conjunction with optical detection~\cite{Bhattacharjee2020a} {\color{revised}in a quasi-unbounded environment in contrast to a pipe-based testbed environment}. That way, it is possible to investigate the influence of ventilation on the spatial-temporal distribution of the cloud.  Furthermore, it is possible to take the influence of obstructions like protection masks into account. {\color{revised} The efficacy of a medical mask is verified.} 
	\item An advanced computer simulation tool, originally developed for efficient simulation of MC in fluid channels~\cite{drees2020efficient}, is {\color{revised} extended. This tool can be} used to estimate transmission of infectious aerosols in different,  {\color{revised}preferably indoor} environments. Parameters feeding this simulation tool are partly obtained by self experiments, among other established sources. 
\end{itemize}

Particle concentrations (``clouds'') are released by respiratory events like breathing, speaking, singing, coughing, and sneezing. The aerosols are not only received by mouth and nose, but also by hands, increasing the effective aperture of the receivers, similar to spatially distributed antennas. The probability of infection is not only determined by peer-to-peer distances (motivating ``social distancing''), but also by ventilation of the environment, movement of the receiver apertures through the cloud and individual immune responses of the receiving humans.

Understanding the impact of different channel configurations leads to implications for infection prevention in epidemiology. This helps in developing guidelines for maintaining safe distances and to take preventive measures in order to reduce the probability of infection.

The remainder of this paper is structured as follows: Section~\ref{sec:aerosol_transport} describes the basic transport of aerosol particles in an air-based environment, mentioning effects like diffusion, evaporation, buoyancy, and settlement that define the trajectory of airborne particles. Section~\ref{sec:scenario} introduces the MC scenario under investigation adapted to the domain of infectious aerosol transmission with regard to the aforementioned channel model and its generalization to multiuser communication. In Section~\ref{sec:measurement}, MC measurements with fluorescent particles are adapted to the field of infectiology, accompanied by simulations for the given scenario, before closing with the conclusion and an outlook on future developments and adjustments from the field of MC in Section~\ref{sec:conclusion}.
\section{Principles of Aerosol Transport\label{sec:aerosol_transport}}

\begin{figure}
	\centering
	\usetikzlibrary{backgrounds,fit}
\makeatletter
\tikzset{
	auto centering/.style={execute at end picture={
			\node[fit=(current bounding box),minimum width=#1-2*\tikz@framexsep,inner sep=0,
			]{};
	}},
	auto centering/.default=\columnwidth,
}
\makeatother

\begin{tikzpicture}[auto centering,background rectangle/.style={fill=blue!20},show background rectangle]
	\usetikzlibrary{arrows}
	\usetikzlibrary{arrows.meta}
	\usetikzlibrary{shapes}
	\usetikzlibrary{calc}
	
	\node [fill=red!50,single arrow,minimum height=12em,minimum width=6em] at (3.8,0) {Emission direction};
	
	\node[circle,minimum width=10em,minimum height=10em] at (0,0) (AirStream) {};
	\shade[shading=radial,inner color=red,outer color=blue!20,opacity=.75] (0,0) circle (6em);
	
	\node[fill,draw,circle,minimum width=10pt] at (0,0) (MainDroplet) {};
	
	\shade[draw,shading=linear,left color=blue!20,right color=red,opacity=.75] (3,-2) rectangle (4.5,-2.2);
	\node[above] at (3,-2) (Tmin) {$T_\text{min}$};
	\node[above] at (4.5,-2) (Tmax) {$T_\text{max}$};
	
	
	\draw[->] (MainDroplet.east) -- ($(MainDroplet.east)+(1em,0)$) node[above] {$F_\text{A}$};
	\draw[->] (MainDroplet.south) -- ($(MainDroplet.south)+(0,-1em)$) node[right] {$F_\text{G}$};
	
	\draw[->] (AirStream.west) -- ($(AirStream.west)+(-1em,0)$) node[above] {$F_\text{D}$};
	\draw[->] (AirStream.north) -- ($(AirStream.north)+(0,1em)$) node[right] {$F_\text{B}$};
\end{tikzpicture}%
	\caption{{\color{revised}Forces acting on aerosols in an air-based environment: bouyancy force $F_\text{B}$ and air drag force $F_\text{D}$ act mainly on the air-stream that is expelled from the mouth or nose, whereas advection force $F_\text{A}$ and gravitational force $F_\text{G}$ act on the aerosols themselves.}
	The temperature gradient is given through the minimum temperature $T_\text{min}$ of the air in the environment and the maximum temperature $T_\text{max}$ of the liquid in the oral cavity.}
	\label{fig:aerosol_forces}
\end{figure}
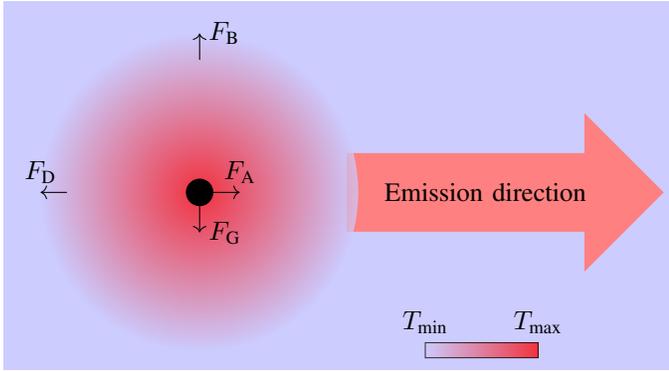

Aerosols are solid or liquid particles found suspended in a gaseous environment with sizes ranging from a few nanometers to several micrometers~\cite{Hinds1999}. In human respiratory events, these particles are expelled with an air stream into the surroundings, where they propagate over large distances, depending on their initial size and velocity. When focusing on the transmission of pathogens in aerosols, sizes in the range of a few micrometers are of special interest, as they are able to carry infectious viral RNA and propagate further into the human respiratory tract, leading to the infection of humans. The type of respiratory event defines the aerosol size and velocity distribution, the overall amount of aerosols, emission angle, and opening angle of the air stream. This multivariate random process of aerosol generation and distribution in the environment defines the risk of infection for other healthy humans.

The transport of aerosols is subject to several effects between the aerosols and the environment, illustrated in Fig.~\ref{fig:aerosol_forces}. The main mode of transportation is aided by the release of an air stream from the mouth or nose, depending on the type of respiratory event. The air stream supports the droplet movement in terms of a drift velocity, often called advection. Through the turbulences generated by the interaction between the released air and the present air in the environment, mainly influenced by air drag, a diffusion process called eddy diffusion acts on the aerosols. This additional diffusion accompanies the mass diffusion already happening at a molecular level. Given the parameters of respiratory events and the sizes of aerosols emitted from these events, the effect of mass diffusion on the movement of the aerosols can be neglected. This results in the eddy diffusion and the advection process being the main modes of transportation for infectious airborne particles.

Due to the temperature difference between the released air and the environment, a density difference results in a buoyancy force acting on the air stream, finally resulting in an upwards movement. On the other hand, gravitation acts on the particles in the air stream, which results in the settlement of particles. The combination of the upwardly directed buoyancy force and the perpendicularly directed gravitational force determines the flight distance of the aerosols. 

Following the same temperature profile as the expelled air stream, the aerosols are subject to an evaporation process resulting in the shrinkage of the droplet diameter{\color{revised}~\cite{Ranz1952}}.  The outer surface of the aerosol changes into gas phase, as the heat is exchanged between the liquid and the environmental air. 

{\color{revised}Larger and heavier droplets are either fragmented into smaller droplets~\cite{Bourouiba2020} or quickly forced to the ground.}
Without additional turbulences and ventilation, aerosols are able to stay suspended in the environment for minutes, resulting in a long lasting risk of infection even after the respiratory event that initially emitted the aerosols.

The combination of the forces acting directly on the aerosols and the forces acting on the air stream, which indirectly influence the aerosols, ultimately lead to the characteristic distributions of aerosols in the ambient air and the associated probabilities of infection.

\section{Scenario under Investigation\label{sec:scenario}}
\begin{figure}
    \centering
    \usetikzlibrary{positioning}
\usetikzlibrary{shapes,arrows}
\usetikzlibrary{3d}

\begin{tikzpicture}[x=0.5cm,y=0.5cm,z=0.3cm,>=stealth]
	\draw[->] (xyz cs:x=0) -- (xyz cs:x=10) node[right] {$x$};
	\draw[->] (xyz cs:y=0) -- (xyz cs:y=4) node[right] {$z$};
	\draw[->] (xyz cs:z=0) -- (xyz cs:z=10) node[above] {$y$};
	\foreach \coo in {0,1,...,9}
	{
		\draw (\coo,-1.5pt) -- (\coo,1.5pt);
		\draw (xyz cs:y=-0.15pt,z=\coo) -- (xyz cs:y=0.15pt,z=\coo);
	}
	\foreach \coo in {0,1,...,3}
	{
		\draw (-1.5pt,\coo) -- (1.5pt,\coo);
	}

	\begin{scope}[canvas is xz plane at y=0]
		\draw[gray!40,step=1.0,thin] (0,0) grid (10,10);
	\end{scope}

	\node[antenna,scale=0.5,fill,color=red] at (xyz cs:x=3,z=2.5) (Tx1) {};
	\node[antenna,scale=0.5,fill,color=blue] at (xyz cs:x=5,z=7) (Rx1) {};
	\node[antenna,scale=0.5,fill,color=blue] at (xyz cs:x=8,z=2) (Rx2) {};
	\node[antenna,scale=0.5,fill,color=blue] at (xyz cs:x=1,z=3.5) (Rx3) {};
	
	\begin{scope}[canvas is xz plane at y=0]
		\path[draw,thick,color=red,->] (1,0) .. controls (2,2) .. (3,2.5);
		\path[draw,thick,color=red,->] (3,2.5) .. controls (4,4) .. (0,7);
		\path[draw,thick,color=blue,->] (0,10) .. controls (1,9) .. (5,7);
		\path[draw,thick,color=blue,->] (5,7) .. controls (8,8) .. (10,10);
		\path[draw,thick,color=blue,->] (10,8) .. controls (9,8) .. (8,2);
		\path[draw,thick,color=blue,->] (8,2) .. controls (8,1) .. (8,0);
		\path[draw,thick,color=blue,->] (0,3) ..controls (0.5,3) .. (1,3.5);
		\path[draw,thick,color=blue,->] (1,3.5) .. controls (5,5) .. (10,3);
	\end{scope}
\end{tikzpicture}
    \caption{Sketch of a multiuser scenario with one infected transmitter {\color{revised}(red)} emitting infectious aerosols into an air-based environment with three healthy receivers {\color{revised}(blue)} at a time instance $t$. {\color{revised}The lines show possible trajectories.}
    }
    \label{fig:scenario_sketch}
\end{figure}
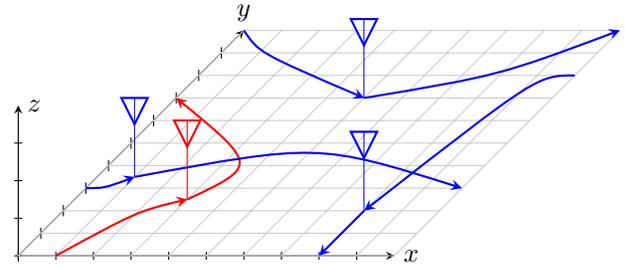

The basic scenario in MC represents a molecular emission device as a transmitter, a propagation channel defined by an impulse response that depends on various environmental parameters, and a receiver interacting with the emitted molecules. Parameters like the distance between the transmitter and the receiver, diffusive behaviour of the molecules in the channel, drift velocities of the environment, and the reception mechanisms at the receiver mainly influence the quality of information exchange in this scenario.

This scenario resembles an elementary peer-to-peer communication scheme with two users. To enable the {\color{revised}translation to} the field of virology, the scenario must be extended to several users who can both send and receive aerosols in a three dimensional (3D) environment. A randomly selected proportion of users are assumed to be infected, emitting infectious particles during their respiratory events. This extended macroscopic MC scenario is shown in Fig.~\ref{fig:scenario_sketch}. All transceivers in the scenario are mobile. This results in a dynamic situation, where the distances between the users are varying.


\subsection{Aerosol Emission}
Partly following~\cite{Khalid2020}, the respiratory events can be separated into two classes: breathing as a continuously occurring event, and speaking, coughing, and sneezing as randomly occurring events with specified parameters. With each event, an air stream containing a cloud of aerosols is emitted, given a size distribution, emission angle distribution, and initial velocity distribution. 
{\color{revised}All these parameters can be interpreted as individual random variables that depend on the type of respiratory event, the infection state of the emitter~\cite{Dhand2020}, and other factors like age and size.}
Depending on the infection status of the emitter, a proportion of the aerosols emitted are assumed to be infectious. {\color{revised}The probability that a droplet contains viruses is closely related to its initial volume~\cite{Stadnytskyi2020}.} These infectious aerosols lead to a risk of infection for the other users present in the environment. 


Having pointed out the duality between coronavirus transmission and air-based macroscopic MC, one item has never been addressed in this concept, namely the modulation scheme and its analogy with data symbols. In good approximation, the modulation scheme can be modeled by respiratory-event-driven higher-order molecular variable-concentration shift keying (MoVCSK)\cite{Bhattacharjee2019}, where the type of molecule {\color{revised}distinguishes infectious from non-infectious particles}. In this way, an infected person can release both infectious and non-infectious particles at the same time. The number of released molecules per unit time depends on the situation.  More molecules are emitted in the case of sneezing and coughing, less molecules are released in the case of breathing and singing. These differences in quantity can be modeled by the different concentration levels of MoVCSK. The different events can be represented by different data symbol patterns. For example, continuous sources such as breathing leads to a burst of data symbols with lower concentration levels, while impulsive sources such as sneezing and coughing lead to individual data symbols with higher concentration levels.

The possible respiratory events are described and modeled in the following paragraphs. Each event triggers the emission of a specific but random distribution of particles into the environment. These initial parameters ultimately determine the range of an aerosol emission and thus the risk of infection for non-infected individuals in the environment.

\subsubsection{Breathing}
The continuous breathing of a human is influenced by the physical load on the body. The breathing pattern at rest of 12-16 breaths per minute{\color{revised}~\cite{Barrett2010}} can increase with higher physical activities, resulting in a higher aerosol emission rate and also larger emission volumes following deeper breaths. 

\subsubsection{Speaking}
Human speaking can be modeled as a two-state Markov process{\color{revised}~\cite{Jaffe1964}}. The states represent the talking and silence periods. This discrete-time model is characterized by the transition probabilities between the states and the retention probabilities for staying in the current state. This represents the fact that a human, currently being in speaking state, might eventually stay in this state for the next time step with a higher probability than switching into the silent state. Whenever a human is speaking, an average aerosol distribution is emitted into the environment. This simplification takes into account differences in languages and phonetic characteristics, as the transition probabilities may be time-varying.

\subsubsection{Coughing and Sneezing}
Coughing and sneezing as randomly and sparsely occurring events can be modeled for each time step in terms of a Bernoulli trial, where the user might eventually emit aerosols in terms of a cough or sneeze, or stay in normal breathing state. The parameter for the resulting Poisson process is influenced by the infection state of the user. The intensity increases for an infected human, which resembles a higher probability for coughing and sneezing, whereas a healthy human might cough or sneeze very rarely. {\color{revised}This concept has been introduced in~\cite{Penrose1996}.}

\subsection{Aerosol Propagation}
The principles of aerosol transport are explained in Section~\ref{sec:aerosol_transport}. The environment under investigation is assumed to be a boundless 3D space filled with air at room temperature with no additional drift or ventilation, aiding the aerosol transport. All aerosol propagation processes are depending on the respiratory event responsible for the emission of the aerosols. This event defines the aerosol size and initial velocity distributions as well as the parameters of the expelled air stream. 

In future research, other room geometries may be studied as well. Additionally, air ventilation and additional turbulences introduced through the movement of the users might be further investigated. 

\subsection{Aerosol Reception}
Aerosol reception is modeled by an effective aperture area, representing the sensitive parts of the body which are mainly responsible for infection via pathogen-laden aerosols. The most prominent area consists of the mouth and nose part of the face. Additional reception can be accomplished by the hands, accounting for smear infections by touching the face. The effective aperture of the subject moves dynamically through the 3D space with the random motion of the human receiver. 

When the effective aperture interferes with suspended aerosols in the air, the detection of aerosols has an impact on the infection probability of the receiver. The viral load in the aerosols and the individual immune response of the receiver defines its infection probability. Each human is equipped with an individual detection threshold, steering infection probabilities at the receiver. Similar to spatially distributed antennas, an antenna gain can be specified for the different areas of the effective aperture, which defines the reception strength of the infectious aerosols and the resulting probability of infection. Since the number of infectious particles reaching the recipient over a certain period of time is relevant for an infection, the receiving process can be modeled as energy detection. It is worth mentioning that in contrast to classical MC, intersymbol interference and multiuser interference do not need to be cancelled.
\section{Measurement Results\label{sec:measurement}}
The real-world measurements reported in this paper are performed with a single emitter to gather information about the propagation behaviour of airborne aerosols from human respiratory events in an air-based environment. This information can be used in the simulation to perform multiuser scenarios. 

Real-world measurements serve as the basis of channel models and computer simulations if an analysis turns out to be too challenging. Measurements are necessary to understand the propagation behaviour of different-sized particles from respiratory events. In experiments, it is also possible to perform more advanced measurements with different types of face protection gear or to investigate the influence of a longer wearing period, to consider air ventilation, and to test arbitrary room geometries. Taking all these points into consideration, the inspiration is drawn from the testbed design used in~\cite{Bhattacharjee2020b} to conduct experiments using fluorescent dyes.
{\color{revised}The industrial sprayer used as a transmitter in~\cite{Bhattacharjee2020a} is replaced by a test person emulating an infected person. As an example of the above mentioned respiratory events, successive coughs are induced. Key parameters obtained from the experiments like particle diameters and distance profile are used in the simulation tool.}
The tube-like structure forming a bounded channel is scaled up to an enclosed environment. The optical-based receiver is treated as an uninfected human, interacting with emitted aerosols and droplets. {\color{revised}The advantage of fluorescent particles becomes apparent, as they stay visible under longwave ultra-violet (UVA) light even after settlement. Commonly used techniques like shadowgraph~\cite{Tang2012} or Schlieren imaging~\cite{Tang2009} do not provide this feature. }

\subsection{Experiments\label{ssec:experiment}}
\begin{table}
	\caption{Parameters and measures of the experiments.}
	\label{tab:experiment_parameters}
	\centering
	\begin{tabular}{ll}
	    \toprule
		Parameter & Value \\
		\midrule
		Emission height & $\SI{1.64}{\meter}$ \\
		Temperature inside mouth & $\SI{36.0}{\celsius}$ \\
		Room temperature & $\SI{20.1}{\celsius}$ \\
		Room length & $\SI{2.34}{\meter}$ \\
		Room width & $\SI{1.70}{\meter}$ \\
		Dye concentration & $\SI{1.54}{\gram\per\liter}$ \\
		Number of test series & 5 \\
		Number of coughs per series & 5 \\
		\bottomrule
	\end{tabular}
\end{table}

The tests are carried out with a test person to obtain sufficient parameters for the simulation environment: 
{\color{revised} About $\SI{10}{\milli\liter}$ of a saturated aqueous solution containing the fluorescent dye Uranine (mixing ratio: $\SI{1.54}{\gram\per\liter}$) is taken in through the mouth and mixed with saliva. Over a period of two minutes the liquid in the mouth cavity is heated to body temperature.} 
At the time of an artificially induced cough, the particles are in the mouth and throat of the test person. As a result of the cough impulse, the particles are partially expelled and flung into the ambient air as airborne aerosols and droplets, following a characteristic diameter distribution. The trajectory of the aerosols is made visible using a \SI{50}{\watt} UVA lamp with a peak emission wavelength of $\SI{396}{\nano\meter}$ corresponding to the excitation wavelength of the fluorescent dye. Uranine responds with an emission wavelength of \SIrange{520}{530}{\nano\meter}, which is clearly distinguishable in the visible light spectrum. Recording is accomplished by a high-speed camera with $\SI{240}{\fps}$ placed orthogonal to the flight direction of the aerosols.

The test person was an approximately sixty-year-old male. The cough impulse was expelled at a height of $\SI{1.64}{\meter}$ in horizontal direction. At the time of the measurements, the test person was completely healthy and had no physical symptoms of disease. The temperature inside the mouth was $\SI{36.0}{\celsius}$ and the Uranine solution was heated to this temperature for a sufficiently long time. Experiments were performed in a fanless dark room. The room had an air temperature of $\SI{20.1}{\celsius}$ and a length of $\SI{2.34}{\meter}$, with $\SI{0.85}{\meter}$ of space on each side orthogonal to the direction of emission up to the adjacent walls. 
{\color{revised} Four test series were carried out without a protective mask and one test series with a mask. Each test series consists of five consecutive, artificially triggered coughs, each about 15 seconds apart. The floor was photographed before each new cough and after the final cough. Between the test series, a new portion of Uranine solution was taken into the mouth and warmed to body temperature, and the floor was cleaned of the residues of the previous test series.}
Key parameters of the experiments are summarized in Table~\ref{tab:experiment_parameters}. The measurement results reported next have been performed without protection mask or face shield, unless mentioned specifically. To {\color{revised} highlight the importance} of a face-mask as an obstruction device to reduce the infectious range of airborne aerosols, {\color{revised} one test series was} taken with a well-fitting medical filtering-face-piece (FFP1) face-mask (\textit{3M Aura 1861+}) according to EN149:2001 FFP1 NR D {\color{revised} and Type IIR EN 14683:2019, having a bacterial filtration efficiency of more than 98~\% and an aerosol filtration percentage of 80~\% minimum.}

\begin{figure}
	\centering
	\begin{subfigure}[c]{0.49\linewidth}
		\includegraphics[width=\textwidth]{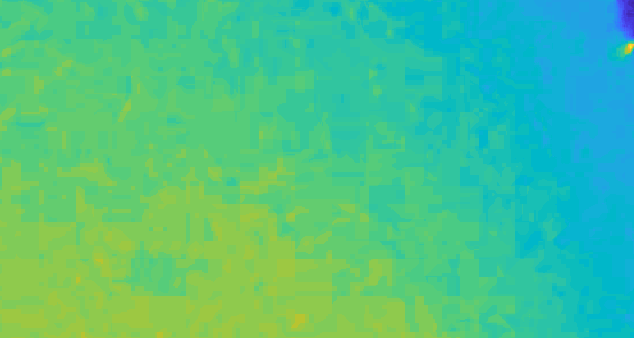}
		\subcaption{$t=t_\text{e}+\SI{0}{\second}$}
	\end{subfigure}
	\begin{subfigure}[c]{0.49\linewidth}
		\includegraphics[width=\textwidth]{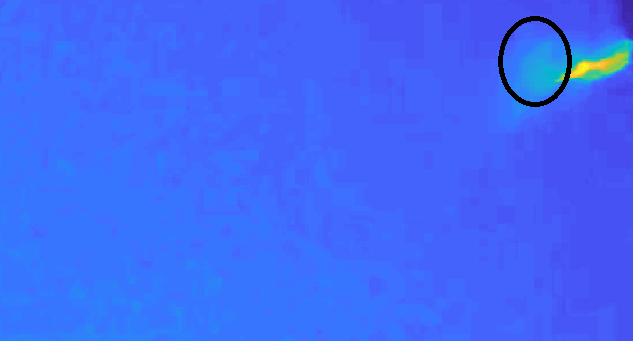}
		\subcaption{$t=t_\text{e}+\SI{0.029}{\second}$}
	\end{subfigure}

	\begin{subfigure}[c]{0.49\linewidth}
		\includegraphics[width=\textwidth]{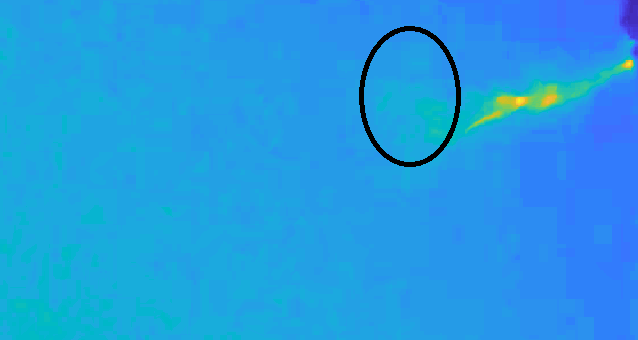}
		\subcaption{$t=t_\text{e}+\SI{0.058}{\second}$}
	\end{subfigure}
	\begin{subfigure}[c]{0.49\linewidth}
		\includegraphics[width=\textwidth]{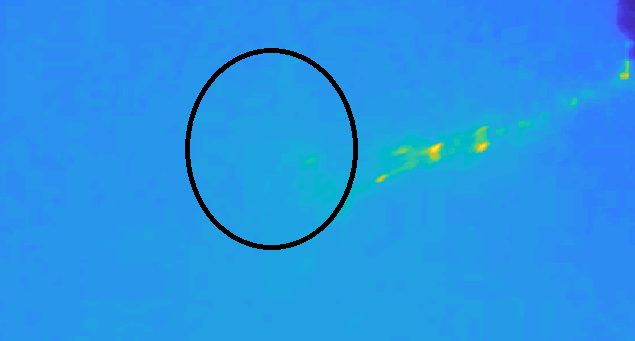}
		\subcaption{$t=t_\text{e}+\SI{0.092}{\second}$}
	\end{subfigure}
	\caption{Fluorescent aerosols expelled through coughing action at emission time $t_\text{e}$. The droplets are visible as luminous dots, aerosols are visible as clouds and highlighted by the circle.}
	\label{fig:video_first_frames}
\end{figure}

Characteristic frames of the video are presented in Fig.~\ref{fig:video_first_frames}. They show the development of aerosols and droplets directly after the cough impulse in the ambient air.  After approximately \SI{30}{\milli\second}, a difference between the droplets in a jet-like formation and the aerosols in a cloud formation is clearly visible. This distinction continues in the following frames and becomes even clearer as the droplets sink to the ground due to the gravitational force, following the trajectory of a horizontal throw. The aerosols in the cloud formation are carried further through the room by the ejected air stream, as explained in Section~\ref{sec:aerosol_transport}. These aerosols are of special interest for infection prevention techniques, mainly because they are able to stay in the environment for a longer time and, thus, lead to longer infection periods. They are also able to travel larger distances due to the air stream and turbulence, thus, causing infections by entering the respiratory tract directly via the inhaled air. The larger particles can be absorbed through the effective aperture of the receptor and can cause smear infections if the receptor infects their own airways through improper use.

With the calibrated video recordings, it is possible to determine the velocity of the aerosol cloud and the droplet jet. These measurements can be used to determine theoretical distances of the different modes of aerosol emission and, thus, leading to implications for prevention techniques with infectious aerosols. The average initial velocity of the aerosol cloud is about $\SI{8}{\meter\per\second}$, whereas the droplets are ejected at about $\SI{5}{\meter\per\second}$. From the recordings, it seems natural to estimate the maximum travel distance $d_\text{max}$ of the droplets by assuming a horizontal throw with a maximum flight distance of
$d_\text{max} = v \, \sqrt{2h_0/g}$, where $v$ is the droplet velocity, $h_0$ the emission height and $g$ the gravitational acceleration. This assumption results in a maximum distance of $d_\text{max}=\SI{2.90}{\meter}$ that the droplets are able to reach. This result is in accordance with the observation that several droplets were hitting the $\SI{2.34}{\meter}$ distant wall, up to a considerable height of $\SI{1.35}{\meter}$ given an initial level of $\SI{1.64}{\meter}$.

\begin{figure}
	\centering
	\includegraphics[width=\linewidth]{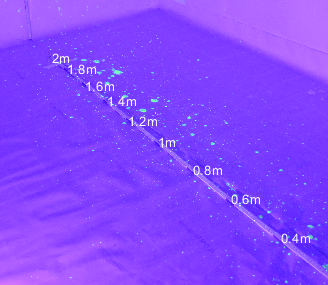}
	\caption{Fluorescent droplets on the floor after five artificially induced coughs from a height of $\SI{1.64}{\meter}$.}
	\label{fig:droplets_floor_distribution}
\end{figure}

As an additional instrument to the video capturing of the cough itself, the floor is photographed after each cough, while it is also illuminated with the UVA lamp to show the fluorescent dots from the settlement on the ground. The resulting photographs are used to determine the trajectories of the droplets, which help in estimating the infectious range as well as the diameter of the droplets. An example image of this process is shown in Fig.~\ref{fig:droplets_floor_distribution} with a ruler for the distance estimation and overall calibration of the image analysis.

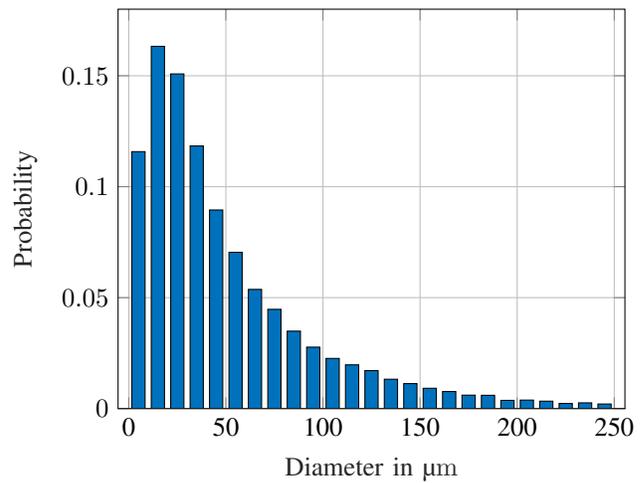
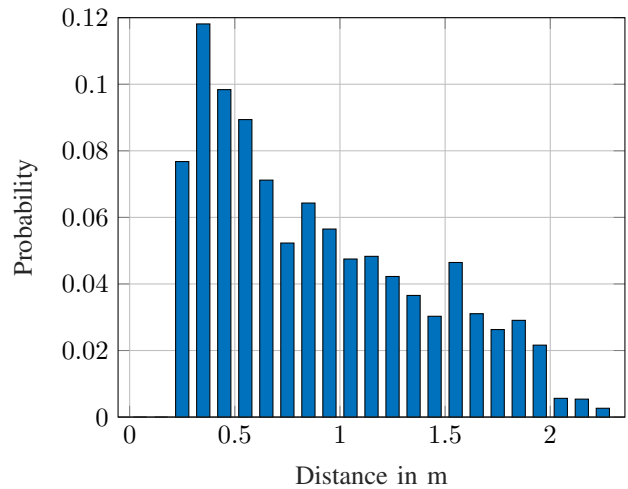
\begin{figure}
	\centering
	\newlength\fwidth
	\setlength\fwidth{0.8\linewidth}
	\begin{subfigure}[c]{\linewidth}
%
%
\definecolor{mycolor1}{rgb}{0.00000,0.44700,0.74100}%
\begin{tikzpicture}

\begin{axis}[%
width=0.951\fwidth,
height=0.75\fwidth,
at={(0\fwidth,0\fwidth)},
scale only axis,
scaled ticks=false,
y tick label style={/pgf/number format/fixed},
bar shift auto,
xmin=-5.49999999999996e-06,
xmax=0.0002555,
xticklabels={0,0,50,100,150,200,250},
xlabel style={font=\color{white!15!black}},
xlabel={Diameter in $\si{\micro\meter}$},
ymin=0,
ymax=0.18,
ylabel style={font=\color{white!15!black}},
ylabel={Probability},
axis background/.style={fill=white},
xmajorgrids,
ymajorgrids
]
\addplot[ybar, bar width=5, fill=mycolor1, draw=black, area legend] table[row sep=crcr] {%
5e-06	0.115804851325441\\
1.5e-05	0.16325288370819\\
2.5e-05	0.15083543841495\\
3.5e-05	0.118361384179931\\
4.5e-05	0.0895090848221079\\
5.5e-05	0.0704263931582311\\
6.5e-05	0.0537176248592385\\
7.5e-05	0.0447393249535868\\
8.5e-05	0.0349088474297714\\
9.5e-05	0.0276957725903156\\
0.000105	0.0226131417962687\\
0.000115	0.0197218248774995\\
0.000125	0.0171348571080744\\
0.000135	0.0132391879964696\\
0.000145	0.0112609185257327\\
0.000155	0.00916090939525824\\
0.000165	0.00770003347840643\\
0.000175	0.00605654807194814\\
0.000185	0.00599567824207931\\
0.000195	0.00371305962199836\\
0.000205	0.00383479928173601\\
0.000215	0.0033478406427854\\
0.000225	0.00231305353501537\\
0.000235	0.00258696776942508\\
0.000245	0.00206957421554007\\
};
\addplot[forget plot, color=white!15!black] table[row sep=crcr] {%
-6.99999999999995e-06	0\\
0.000257	0\\
};
\end{axis}
\end{tikzpicture}%
		\caption{Distribution of droplet diameters.}
		\label{fig:aerosol_diameter_distribution}
	\end{subfigure}
	
	\begin{subfigure}[c]{\linewidth}
%
%
\definecolor{mycolor1}{rgb}{0.00000,0.44700,0.74100}%
\begin{tikzpicture}

\begin{axis}[%
width=0.951\fwidth,
height=0.75\fwidth,
at={(0\fwidth,0\fwidth)},
scale only axis,
scaled ticks=false,
y tick label style={/pgf/number format/fixed},
bar shift auto,
xmin=-0.0549999999999996,
xmax=2.355,
xlabel style={font=\color{white!15!black}},
xlabel={Distance in m},
ymin=0,
ymax=0.12,
ylabel style={font=\color{white!15!black}},
ylabel={Probability},
axis background/.style={fill=white},
xmajorgrids,
ymajorgrids
]
\addplot[ybar, bar width=5, fill=mycolor1, draw=black, area legend] table[row sep=crcr] {%
0.05	0\\
0.15	0\\
0.25	0.0767694234094568\\
0.35	0.118120565261463\\
0.45	0.0983841154373621\\
0.55	0.0893804782064278\\
0.65	0.0711943235346729\\
0.75	0.0522926480233737\\
0.85	0.0643074354540576\\
0.95	0.0564963329556973\\
1.05	0.047492695724763\\
1.15	0.0482976566692505\\
1.25	0.042245542901437\\
1.35	0.0365511895533957\\
1.45	0.0302903822073818\\
1.55	0.0464492278337607\\
1.65	0.0310655297835549\\
1.75	0.0262953908532586\\
1.85	0.0290680341064934\\
1.95	0.0216146920279053\\
2.05	0.00563472661141256\\
2.15	0.00539621966489774\\
2.25	0.00265338977997734\\
};
\addplot[forget plot, color=white!15!black] table[row sep=crcr] {%
-0.0699999999999996	0\\
2.37	0\\
};
\end{axis}
\end{tikzpicture}%
		\caption{Distribution of droplet flight distances.}
		\label{fig:aerosol_distance_distribution}
	\end{subfigure}
	
	\caption{Combined distributions for a test block of five consecutive coughs by the test person.}
\end{figure}

{\color{revised}To estimate the droplet diameter distribution shown in Fig.~\ref{fig:aerosol_diameter_distribution}, the photographs are analyzed with an automated approach employing Matlab.
A tape measure laid out on the floor enables a calculation of the size of the droplets on the floor given the recorded pixels.
A calibration measurement obtained by an industrial sprayer with a defined droplet radius of $\SI{3.3}{\micro\meter}$ is then used to relate the radius of circles on the floor to the actual droplet sizes in the air. 
This processing is accompanied by an analysis of distance measurements, as shown in Fig.~\ref{fig:aerosol_distance_distribution}.}

The diameter distribution shows an observed maximum in the range of \SIrange{20}{30}{\micro\meter} with most droplets being under $\SI{100}{\micro\meter}$ in diameter. It is assumed that virus-bearing aerosols are in the range of \SIrange{1}{100}{\micro\meter}, so the given results are relevant for this regime. The distance distribution on the other hand shows droplets up to a distance greater than $\SI{2}{\meter}$, which fits the previous calculations for droplets, given additional effects of air drag and turbulences, reducing the maximum travel distance. An additional factor for limiting the range of the aerosols is the emission angle from the respiratory event: Most of the measurements showed an emission cone pointing downwards, which already accelerated the heavier particles towards the ground. 

The measurements with an FFP1 mask showed no visible fluorescence in the area in front of the test person. After the test interval, the mask was fluorescent, indicating that it absorbed the aerosols. 

\subsection{Simulations}
\begin{table}
    \centering
    \caption{Parameters of the simulation.}
    \label{tab:simulation_parameters}
    \begin{tabular}{ll}
        \toprule
        Parameter & Value \\
        \midrule
        Air density & $\SI{1.2041}{\kilo\gram\per\cubic\meter}$ \\
        Air viscosity & $\SI{18.13e-6}{\newton\second\per\square\meter}$ \\
        Particle mass density & $\SI{997}{\kilo\gram\per\cubic\meter}$ \\
        Emission height & $\SI{1.64}{\meter}$ \\
        Emission opening angle & $\mathcal{N}(0, \SI{6.25}{\degree})$ \\
        Number of particles & 5000 \\
        \bottomrule
    \end{tabular}
\end{table}

For simulating the transmission of particles, the Pogona simulator\textcolor{lukas}{\footnote{\texttt{https://www2.tkn.tu-berlin.de/software/pogona/}}} was used, which was previously developed by the authors for macroscopic MC~\cite{drees2020efficient}.
In~\cite{Bhattacharjee2020a}, this simulator was extended for the propagation of water droplets through air, based on an initial velocity vector, without taking into consideration any other forces acting on the particles.
This tool has now been extended to take different forces explained in Section~\ref{sec:aerosol_transport} into account.

Simulations are performed in a similar 3D environment as the real-world measurements.
The room has a boundless structure and is filled with air at a temperature of $\SI{20}{\celsius}$.
The aerosols are emitted from a source at an initial height of $\SI{1.64}{\meter}$ with an initial emission vector parallel to the ground.
The initial velocity of each particle is sampled randomly according to the empirical cumulative distribution function of velocities derived from the experimentally observed travel distances as explained in Section~\ref{ssec:experiment}, assuming a frictionless horizontal throw.
Similarly, the droplet radii are distributed according to the measured droplet sizes.
In order to speed up computations for this proof of concept, however, particle diameters below \SI{50}{\micro\meter} are omitted.
This is due to the fact that adaptively chosen time steps become extremely small in order to accurately update the velocity of very light-weight particles.
For each particle traveling through the air, the forces are simulated which are detailed in Section \ref{sec:aerosol_transport}, i.e., gravity, drag, and buoyancy.


\begin{figure}
	\centering
	\includegraphics[width=\linewidth]{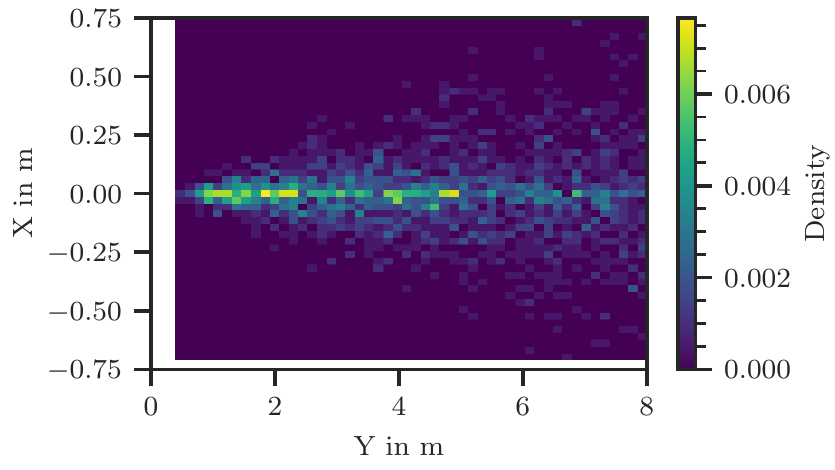}
	\caption{
	    Spatial distribution of simulated particles falling on the floor.
	    The emitter is placed in the origin, facing in positive Y-direction.
	}
	\label{fig:simulation_heatmap}
\end{figure}

In Fig.~\ref{fig:simulation_heatmap}, the distribution shows the accumulation of simulated particles on the floor.
In comparison with Fig.~\ref{fig:aerosol_distance_distribution}, it is seen in both cases that a large portion of particles accumulates on the floor between a distance of \SI{0.5}{\meter} and \SI{2}{\meter}.
In the simulation environment, this region gets extended to a distance of at least \SI{5}{\meter}.
Possible reasons for this may include not simulating the effects of evaporation, simplified assumptions about the initial velocities of particles, or the challenge of coughing in a direction horizontal to the ground in the experiment.
At the same time, it is observed that individual simulated particles travel significantly farther than \SI{5}{\meter}, which is in line with the observation above that droplets can be found at a height of over \SI{1}{\meter} on the wall at the other side of the room.

\subsection{\color{revised} Relation to Communication Theory and Networking\label{sec:relations}}
\begin{table*}[t]
\caption{\color{revised} Relationship between conventional performance metrics and infectious particle transmission.}
\centering
\color{revised}
\begin{tabular}{ll}
\toprule
Traditional performance metric & Infectious particle transmission \\
\midrule
Received power (signal-to-noise ratio)  & Heat map \\
Communication range                     & Infection range \\
Beamforming                             & Shapes of aerosol clouds \\
Signal blocking / shadowing             & Masks, protective screens, and other obstacles \\
Retransmission                          & Consecutive respiratory events \\
\midrule
Outage, bit/frame error rate            & Probability of infection \\
Connectivity                            & Chain of infection, reproduction number \\
Throughput, data rate                   & Viral load \\
Latency (jitter)                        & Incubation time \\
Number of communication channels        & Number of contacts \\
\bottomrule
\end{tabular}
\label{table:relation}
\end{table*}

{\color{revised}
The duality elaborated in Section~\ref{sec:scenario} makes it possible to relate selected results and conclusions of the experiments and simulations to traditional performance metrics in communication theory and networking: 
\begin{enumerate}
    \item The spatially distributed received power (or the signal-to-noise ratio, respectively) relates to the heat map in Fig.~\ref{fig:simulation_heatmap}.
    \item The communication range corresponds to the range of infection. 
    The distribution of droplet flight distances is depicted in Fig.~\ref{fig:aerosol_distance_distribution}. The maximum range, $d_\mathrm{max}$, has been confirmed analytically.
    \item The shapes of the emitted aerosol clouds depend on the type of respiratory event and are analogous to beamforming, i.e., to directional transmission of information. 
    These shapes are considered in the simulation tool. 
    \item Signal blocking/shadowing is linked to masks, protective screens, and other obstacles.
    \item Retransmission is related to consecutive respiratory events. 
\end{enumerate}

These and additional performance metrics that are not further evaluated in this contribution (such as probability of infection, chain of infection, viral load, incubation time) are depicted in Table~\ref{table:relation}. 
}

\section{Conclusion\label{sec:conclusion}}
{\color{revised} This paper exploits the duality between a viral infection process and macroscopic air-based molecular communication.  
It connects an engineering perspective with the medical domain of virus-laden airborne particles as disease spreading mechanism.
Novel contributions include the following aspects:
\begin{itemize}
    \item The transmission of infectious and non-infectious aerosols and droplets from human airways expelled through respiratory events is interpreted as a multiuser MC scenario. Infection chains are the root of reproduction. 
    \item Respiratory events like breathing, speaking, coughing, or sneezing are modeled by respiratory-event-driven molecular concentration shift keying. 
    \item Channel modeling is aided by human experiments employing a fluorescent dye.  The efficacy of a medical mask is verified.
    \item  The advanced MC Pogona simulation tool is extended to emulate aerosol transmission in different environments.
    Parameters feeding this tool are partly obtained by own experimental results. 
\end{itemize}
The contribution paves the way to new opportunities and tools for analysis and prevention of infectious diseases caused by airborne pathogen transmission.

An optical particle detection technique previously employed in the area of air-based macroscopic MC is used as a basis to model the different respiratory events. Video recordings and existing knowledge about channel modeling from MC are combined to make more accurate predictions about the dispersion of aerosols in the ambient air using an advanced simulation environment. Possibly infectious particles are made visible using a fluorescent dye.}

Real-world experimental measurements have shown that a bimodal distribution of particles is emitted when coughing, caused by aerosols and droplets, respectively. Aerosols are transported by the expelled air and reach an average initial velocity of about $\SI{8}{\meter\per\second}$. As a result, these small particles can travel significant distances and cause infections even outside the previously assumed safe distances. Moreover, aerosols do not follow a parabolic path and may remain in the ambient air for a notable period of time, as they are transported with the expelled air. This results in a long time of risk of infection. On the other side, droplets {\color{revised} are either fragmented into smaller particles, or they} follow a parabolic path towards the ground. Their ejection speed of around $\SI{5}{\meter\per\second}$ is also relatively high, which leads to long flight distances of $\SI{2}{\meter}$ and beyond. These experiments are reproduced using the Pogona simulator, where a spatial distribution of simulated particles emitted via coughing is generated. It is seen that the results are very much in tune with the real-world measurements as a significant number of particles settle on the floor between a distance of \SI{0.5}{\meter} and \SI{2}{\meter} from the source, with a good portion of particles reaching distances even beyond. Due to the fact that droplets (earlier) and aerosols (later) concentrate on the floor, increased attention should be paid to shoe cleaning.     

Measurements with a medical face mask have shown that this preventive technique is able to reduce the amount of aerosols and droplets emitted from respiratory events considerably. 
{\color{revised} This suggests that the remaining particle concentration outside the mask is below the resolution of the used measuring devices.} 

In future research, the concept introduced in Section~\ref{sec:scenario} may be investigated more deeply, featuring dynamic multiuser situations. Stochastic techniques may also be used to define safe areas in dynamic environments to stop a pandemic infection. Other preventive techniques like ventilation, humidifiers, cloth face masks, or other obstacles may be investigated in terms of leakage after a longer wearing period, to better suit real-world problems.


%


\section*{Acknowledgment}
Reported research was supported in part by the project MAMOKO funded by the German Federal Ministry of Education and Research (BMBF) under grant numbers 16KIS0915 and 16KIS0917.

\ifCLASSOPTIONcaptionsoff
  \newpage
\fi



\bibliographystyle{IEEEtran}
\bibliography{IEEEabrv,literature}
\end{document}